\documentstyle[proceedings,amsmath,amsfonts,amssymb,numreferences]{crckapb}

\begin{opening}
\title{The CKP hierarchy and the WDVV prepotential}
\author{Henrik Aratyn}
\institute{
Department of Physics,\\
University of Illinois at Chicago,\\
845 W. Taylor St.,\\
Chicago, IL 60607-7059\\
e-mail: aratyn@uic.edu}
\author{Johan van de Leur}
\institute{
Mathematical Institute,\\
University of Utrecht,\\
P.O. Box 80010, 3508 TA Utrecht,\\
The Netherlands\\
e-mail: vdleur@math.uu.nl}

\end{opening}
\begin{document}
\maketitle

\section{The WDVV prepotential}
In terms of the so-called flat coordinates
$x^1, x^2, {\ldots} ,x^n$
a solution to the
Witten--Dijkgraaf--Verlinde--Verlinde (WDVV) equations \cite{W1}, \cite{DVV}
is given by a prepotential
$ F (x^1, x^2, {\ldots} ,x^n)$
which satisfies the associativity relations:
\begin{equation}
\sum_{\delta,\gamma=1}^n \frac{\partial^3 F (x) }{\partial x^\alpha \partial
x^\beta \partial x^\delta }\eta^{\delta \gamma}
\frac{\partial^3 F (x) }{\partial x^\gamma \partial x^\omega \partial x^\rho }
=\sum_{\delta, \gamma=1}^n
\frac{\partial^3 F (x) }{\partial x^\alpha \partial x^\omega \partial x^\delta
}
\eta^{\delta \gamma}
\frac{\partial^3 F (x) }{\partial x^\gamma \partial x^\beta \partial x^\rho }
\label{associativity}
\end{equation}
together with
a quasi-homogeneity condition :
\begin{equation}
\sum_{\alpha=1}^n \left( 1+\mu_1-\mu_\alpha \right) x^\alpha \frac{\partial
F}{\partial x^\alpha}
= (3-d) F + \, {\rm quadratic~terms}\, .
\label{quasi}
\end{equation}
where $\mu_i, i=1,{\ldots} ,n$ and $d$ are constants .

Furthermore, expression
\begin{equation}
\frac{\partial^3 F (x) }{\partial x^\alpha \partial x^\beta \partial x^1 }
= \eta_{\alpha\beta} 
\label{eta}
\end{equation}
defines a constant non-degenerate metric:
$g = \sum_{\alpha, \beta =1}^n \eta_{\alpha \beta} dx^\alpha dx^\beta$.

As shown by Dubrovin (e.g. in reference \cite{Du2})
there is an alternative description of the metric in terms of a special class
of orthogonal curvilinear coordinates $u_1, {\ldots} ,u_n$
\begin{equation}
g = \sum_{\alpha \beta =1}^n \eta_{\alpha \beta} dx^\alpha dx^\beta=
\sum_{i=1}^n h_i^2 ({ u}) (d u_i)^2
\label{lam}
\end{equation}
called canonical coordinates.
These coordinates allow to reformulate the problem in terms of the
Darboux-Egoroff
metric systems and corresponding Darboux-Egoroff equations and
their solutions.
In the Darboux-Egoroff metric the Lam\'e coefficients $ h_i^2 ({ u})$
are gradients of some potential and this  ensures that the
so-called ``rotation coefficients''
\begin{equation}
\beta_{ij} = \frac{1}{h_j} \frac{\partial h_i}{\partial u_j}, \;\;  i\ne j,
\;\; 1\le i,j\le n ,
\label{rotco}
\end{equation}
are symmetric $\beta_{ij} = \beta_{ji}$.
The Darboux-Egoroff equations for the rotation coefficients are:
\begin{equation}
    \frac{\partial }{ \partial u_k} \beta_{ij} = \beta_{ik} \beta_{kj},
\;\; \;\mbox{distinct}\;\; i,j,k
\label{betas-comp}
\end{equation}
\begin{equation}
\sum_{k=1}^{n} \frac{\partial }{ \partial u_k} \beta_{ij}     =0,
\;\; i \ne j \, .
\label{ionb}
\end{equation}
In addition to these equations one also assumes the conformal condition :
\begin{equation}
\sum_{k=1}^nu_k\frac{\partial }{ \partial u_k} \beta_{ij} = -\beta_{ij}\, .
\label{betas-deg}
\end{equation}
The Darboux-Egoroff equations (\ref{betas-comp})-(\ref{ionb})
appear as compatibility equations of a linear system :
\begin{eqnarray}
\frac{\partial \Phi_{ij} (u,z) }{\partial u_k} &=& \beta_{ik} (u) \Phi_{kj}
(u,z)
\;\;\; i \ne k\label{delinsa} \\
\sum_{k=1}^n \frac{\partial \Phi_{ij} (u,z) }{\partial u_k} &=&
z \Phi_{ij}  (u,z)
\label{delinsb}
\end{eqnarray}
Define the $n \times n$ matrices $\Phi =(\Phi_{ij})_{1\le i,j\le n}$,
$B =(\beta_{ij})_{1\le i,j\le n}$ and
$ V_i = \left\lbrack B \, ,\, E_{ii} \right\rbrack$, where
$(E_{ij})_{k\ell}=\delta_{ik}\delta_{j\ell}$.
Then the linear system (\ref{delinsa})-(\ref{delinsb})
acquires the following form :
\begin{eqnarray}
\frac{\partial \Phi (u,z) }{\partial u_i} &=&
\left( z E_{ii} + V_i (u) \right)\Phi (u,z), \quad
i=1,{\ldots} ,n \;, \label{delinsma} \\
\sum_{k=1}^n \frac{\partial \Phi (u,z) }{\partial u_k} &=&
z \Phi  (u,z) \, .
\label{delinsmb}
\end{eqnarray}
Let, furthemore $\Phi (u,z)$ have a power series expansion
\begin{equation}
\Phi (u,z) =  \sum_{j=0}^{\infty} z^j
 \Phi^{(j)} (u)=
\Phi^{(0)}(u) + z \Phi^{(1)}(u) + z^2 \Phi^{(2)}(u)+ \cdots
\label{phiexpc}
\end{equation}
and satisfy the ``twisting'' condition ;
\begin{equation}
\Phi (u,z) \eta^{-1} \Phi^T (u,-z)= I
\label{twist}
\end{equation}
Equation (\ref{delinsma}) implies
$ \sum_{k=1}^n u_k {\partial \Phi (u,z) }/{\partial u_k}=
(z U + [B,U]) \Phi (u,z)$, with $U = \sum_{k=1}^n u_k E_{kk}$.
For a matrix $ [B,U] $ which is diagonalizable
the conformal condition (\ref{betas-deg}) leads to
\begin{equation}
\sum_{k=1}^nu_k\frac{\partial \Phi (u,z) }{ \partial u_k}=
z \frac{\partial \Phi (u,z) }{ \partial z} + \Phi (u,z) \mu, \quad \;\;
\mu = {\rm {diag}}  (\mu_1, {\ldots} , \mu_n)
\label{confmat}
\end{equation}
where $\mu$ is a constant diagonal matrix
 obtained by a similarity transformation
from the matrix $[B,U]$. The constant diagonal elements $\mu_i$
entered the quasi-homogeneity condition (\ref{quasi}).

Define
\begin{equation}
\phi_\alpha  (u,z) \equiv
\sum_{\beta=1}^n \Phi^{(0)}_{\beta 1}(u) \Phi_{\beta \alpha} (u,z)
= \phi_\alpha^{(0)} (u)  + z \phi_\alpha^{(1)} (u) + z^2
\phi_\alpha^{(2)}  (u) +  z^3 \phi_\alpha^{(3)} (u) +{\ldots}
\label{xiexp}
\end{equation}
then, in terms of the flat coordinates $x^1, {\ldots} ,x^n$
\begin{equation}
\phi_\alpha^{(1)} (u) = \sum_{\beta=1}^n  \eta_{\alpha \beta} x^\beta  (u)
\label{xione}
\end{equation}
and the prepotential
is given by a closed expression (see e.g. \cite{AKrV} or \cite{AGLZ}):
\begin{equation}
F = - \frac{1}{2} \phi_{1}^{(3)} (u) +
\frac{1}{2} \sum_{\delta=1}^n x^\delta (u)  \phi_{\delta}^{(2)} (u) .
\label{fep}
\end{equation}

\section{The CKP hierarchy}

The CKP hierarchy \cite{Date81} can be obtained as a reduction of the KP
hierarchy,
\begin{equation}
\label{1}
\frac{\partial }{ \partial t_n} { L}  = \lbrack ({
L}^n)_{+}\, , \,  { L}  \rbrack ,\qquad\mbox{for }
{ L} =L(t,\partial)= {\partial_x} + \ell^{(-1)}{\partial_x}^{-1}+
\ell^{(-2)}{\partial_x}^{-2}+\cdots,
\end{equation}
where $x=t_1$, by assuming the extra condition
\begin{equation}
\label{2}
L^*=-L.
\end{equation}
By taking the adjoint, i.e.,$\ ^*$  of (\ref{1}),
one sees that $\frac{\partial L}{\partial t_n}=0$ for $n$ even. Date, Jimbo,
Kashiwara and Miwa \cite{Date81}, \cite{JM} construct such $L$'s from certain
special KP wave functions  $\psi(t,z)=P(t,z)e^{\sum_it_iz^i}$
(recall $L(t,\partial)=P(t,\partial)\partial  P(t,\partial)^{-1}$), where one
then puts all even times $t_n$ equal to 0.
Recall that a KP wave function satisfies
\begin{equation}
\label{2a}
L\psi(t,z)=z\psi(t,z),\qquad
\frac{\partial \psi(t,z)}{ \partial t_n}= (
L^n)_{+} \psi(t,z),
\end{equation}
and
\begin{equation}
\label{2b}
Res\, \psi(t,z)\psi^*(s,z)=0.
\end{equation}
The special wave functions which lead to an $L$ that satisfies (\ref{2})
satisfy
\begin{equation}
\label{3}
\psi^*(t,z)=\psi(\tilde t,-z),
\qquad \mbox{where }\quad \tilde t_i=(-)^{i+1}t_i.
\end{equation}
We call such a $\psi$ a  CKP wave function. Note that this implies that
$L(t,\partial)^*=-L(\tilde t,\partial)$
and that
\[
Res \,\psi(t,z)\psi(\tilde s,-z)=0.
\]
One can put all even times equal to 0, but we will not do that here.

The CKP wave functions correspond to very special points in the
Sato Grassmannian, which consists of all linear spaces
\[
W\subset
H_+\oplus H_-={\mathbb C}[z]\oplus z^{-1}{\mathbb C}[[z^{-1}]],
\]
such that the projection on $H_+$ has finite index. Namely,
$W$ corresponds to a CKP wave function if for any
$f(z),g(z)\in W$ one has $Res\, f(z)g(-z)=0$. The corresponding
CKP tau functions satisfy $\tau(\tilde t)=\tau(t)$.

We will now generalize this to the multi-component case and show that a
CKP reduction of the multi-component KP hierarchy gives the Darboux-Egoroff
system. The $n$ component
KP hierarchy  \cite{ANP99}, \cite{KvdL} consists of the equations in
$t_j^{(i)}$, $1\le i\le n$,
$j=1,2,\ldots$
\begin{equation}
\label{4}
\frac{\partial }{ \partial t_j^{(i)}} { L}  = \lbrack ({
L}^jC_i)_{+}\, , \,  { L}  \rbrack ,\qquad
\frac{\partial }{ \partial t_j^{(i)}} { C_k}  = \lbrack ({
L}^jC_i)_{+}\, , \,  { C_k}  \rbrack ,
\end{equation}
for the $n\times n$-matrix pseudo-differential operators
\begin{equation}
{ L} = {\partial_x} + L^{(-1)}{\partial_x}^{-1}+
L^{(-2)}{\partial_x}^{-2}+\cdots,\qquad C_i=E_{ii}+C_i^{(-1)}{\partial_x}^{-1}+
C_i^{(-2)}{\partial_x}^{-2}+\cdots,
\end{equation}
$1\le i\le n$, where
$x=t_1^{(1)}+t_1^{(2)}+\cdots+t_1^{(n)}$.
The corresponding wave function has the form
\[
\Psi(t,z)=P(t,z)\exp \left(\sum_{i=1}^n\sum_{j=1}^\infty
t_j^{(i)}z^j E_{ii}\right),\quad\mbox{where }
P(t,z)=I+P^{(-1)}(t)z^{-1}+\cdots,
\]
and satisfies
\begin{equation}
\label{5}
L\Psi(t,z)=z\Psi(t,z),\quad C_i\Psi(t,z)=\Psi(t,z)E_{ii},\quad
\frac{\partial \Psi(t,z)}{ \partial t_j^{(i)}}= (
L^jC_i)_{+} \Psi(t,z)
\end{equation}
and
\[
Res\, \Psi(t,z)\Psi^*(s,z)^T=0.
\]
{}From this we deduce that $L=P(t,\partial_x )\partial_xP(t,\partial_x )^{-1}$
and
$C_i=P(t,\partial_x )E_{ii}P(t,\partial_x )^{-1}$.
Using this, the simplest equations in (\ref{5}) are
\begin{equation}
\label{V}
\frac{\partial \Psi(t,z)}{ \partial t_1^{(i)}}= (zE_{ii}+V_i(t))
\Psi(t,z),
\end{equation}
where $V_i(t)=[B(t), E_{ii}]$ with $B(t)=P^{(-1)}(t)$.
In terms of the matrix coefficients $\beta_{ij}$ of $B$ we obtain
(\ref{betas-comp}) for  $u_i=t^{(i)}_1$.

The Sato Grassmannian becomes vector valued, i.e.,
\[
H_+\oplus H_-=({\mathbb C}[z])^n\oplus z^{-1}({\mathbb C}[[z^{-1}]])^n.
\]

The same restriction as in the 1-component case (\ref{3}), viz.,
\[
\Psi(t,z)=\Psi^*(\tilde t,-z),
\qquad \mbox{where }\quad \tilde t_n^{(i)}=(-)^{n+1}t_n^{(i)}.
\]
leads to
$L^*(\tilde t)=-L(t)$, $C_i^*(\tilde t)=C_i(t)$ and
\begin{equation}
\label{A2}
Res\, \Psi(t,z)\Psi(\tilde s,-z)^T=0,
\end{equation}
which we call the multi-component CKP hierarchy.
But more importantly, it  also gives the restriction
\begin{equation}
\label{DB2}
\beta_{ij}(t)=\beta_{ji}(\tilde t).
\end{equation}
Such CKP wave functions correspond to points $W$ in the Grassmannian for which
\[
Res\, f(z)^Tg(-z)=Res\, \sum_{i=1}^n f_i(z)g_i(-z)=0
\]
for any $f(z)=(f_1(z),f_2(z),\ldots, f_n(z))^T,\
g(z)=(g_1(z),g_2(z),\ldots, g_n(z))^T\in W$.

If we finally assume that $L=\partial_x$, then $\Psi$, $W$
also satisfy
\begin{equation}
\label{A1}
\frac{\partial \Psi(t,z)}{\partial x}=\sum_{i=1}^n\frac{\partial
\Psi(t,z)}{\partial t_1^{(i)}}=z\Psi(t,z),\qquad zW\subset W
\end{equation}
and  thus $\beta_{ij}$ satisfies (\ref{ionb}) for $u_i=t_1^{(i)}$. Now
differentiating (\ref{A2}) $n$ times to $x$  for $n=0,1,2,\ldots$
and applying (\ref{A1}) leads to
\[
\Psi(t,z)\Psi(\tilde t,-z)^T=I.
\]
These special  points in the Grassmannian can all be constructed as follows
\cite{vdL1}.
Let $G(z)$ be an
element in  $GL_n({\mathbb C}[z,z^{-1}])$ that satisfies
\begin{equation}
\label{twist2}
G(z)G(-z)^T=1,
\end{equation}
then $W=G(z)H_+$. Clearly, any two $f(z),
g(z)\in W$ can be written as $f(z)=G(z)a(z),\ g(z)=G(z)b(z)$ with
$a(z), b(z)\in
H_+$, then $
zf(z)=zG(z)a(z)=G(z)za(z)\in W$, since $za(z)\in H_+$. Moreover,
\[
Res\, f(z)^Tg(-z)=Res\, a(z)^TG(z)^TG(-z)b(-z)=Res\, a(z)^Tb(-z)=0.
\]
If we define $M(t,z)=\Psi(t,z)G(z)$, then one can prove  \cite{AL1},
\cite{vdL1} that
\[
M(t,z)=M^{(0)}(t)+M^{(1)}(t)z+M^{(2)}(t)z^2+\cdots
\]
We want to change $M(t,z)$ a bit more. However, we only want to do that for
 very special elements in this twisted loop
group, i.e., to certain points of the Grassmannian that have a basis of
homogeneous elements in $z$.
Let $n=2m$ or $n=2m+1$, choose non-negative integers $\mu_i$
for $1\le i\le m$ and define
$\mu_{n+1-j}=-\mu_j$ and let $\mu_{m+1}=0$ if $n$ is odd.
Then take $G(z)$ of the form
\[
G(z)=N(z)S^{-1}=Nz^{-\mu}S^{-1},\qquad\mbox{where}\quad
\mu=\mbox{diag}({\mu_1},{\mu_2}, \ldots,{\mu_n})
\]
and $N=(n_{ij})_{1\le i,j \le n}$ a constant matrix that satisfies
\begin{equation}
\label{N}
N^TN=\sum_{j=1}^n (-1)^{\mu_j}E_{j,n+1-j}
\end{equation}
and
\[
S=\delta_{n,2m+1}E_{m+1,m+1}+\sum_{j=1}^m \frac{1}{\sqrt 2}
\left( E_{jj}+iE_{n+1-j,j}+
E_{j,n+1-j}-iE_{n+1-j,n+1-j}\right).
\]
Then \cite{AL1}
\[
\sum_{i=1}^n\sum_{j=1}^\infty jt_j^{(i)}\frac{\partial \Psi(t,z)}{\partial
t_j^{(i)}}=z\frac{\partial \Psi(t,z)}{\partial
z},
\]
from which one deduces that
\begin{equation}
\label{conf-b}
\sum_{i=1}^n\sum_{j=1}^\infty jt_j^{(i)}\frac{\partial \beta_{ij}}{\partial
t_j^{(i)}}=-\beta_{ij}.
\end{equation}

Define $\eta=(\eta_{ij})_{1\le i,j\le n}=S^TS=\sum_{i=1}^n E_{i,n+1-i}$ and
denote by
$
\Phi(t,z)=M(t,z)S=\Psi(t,z)G(z)S=\Psi(t,z)N(z)$, then $\Phi(t,z)$ satisfies the
following
relations:
\[
\begin{aligned}
\Phi(t,z)=&\Phi^{(0)}(t)+\Phi^{(1)}(t)z+\Phi^{(2)}(t)z^2+\cdots\\[2mm]
\Phi(t,z)\eta^{-1}\Phi(t,-z)^T=&I\\[2mm]
\frac{\partial \Phi(t,z)}{ \partial t_1^{(i)}}=&
(zE_{ii}+V_i(t))\Phi(t,z)\\[2mm]
\sum_{i=1}^n\frac{\partial \Phi(t,z)}{ \partial t_1^{(i)}}=&z\Phi(t,z),\\[2mm]
\sum_{i=1}^n\sum_{j=1}^\infty jt_j^{(i)}\frac{\partial \Phi(t,z)}{\partial
t_j^{(i)}}=&z\frac{\partial \Phi(t,z)}{\partial
z}+\Phi(t,z) \mu.
\end{aligned}
\]

We next put $t_j^{(i)}=0$ for all $i$ and all $j>1$
and $u_i=t^{(i)}_1$, then we obtain the situation of Section 1.
Define $\phi_\alpha(u,z)$ as in (\ref{xiexp}),
then
$
\phi_\alpha^{(1)}(u)=\sum_{\gamma=1}^n \eta_{\alpha\gamma}x^\gamma(u)
$
and the function
$
F(u)$ given by (\ref{fep})
satisfies the WDVV equations.

\section{An example}

We will now give an example of this construction, viz., the case that $n=3$
(for simplicity) and
$\mu_1=-\mu_3=2$ and $\mu_2=0$. Hence, the point of the Grassmannian is given
by
\[
N(z)H_+= N\begin{pmatrix} z^{-2}&0&0\\ 0&1&0\\ 0&0&z^{2}
\end{pmatrix}H_+.
\]
More precise, let $n_i=(n_{1i},n_{2i},n_{3i})^T$ and $e_1=(1,0,0)^T$,
$e_2=(0,1,0)^T$ and $e_3=(0,0,1)^T$, then this point of the
Grassmannian has as basis
\[
n_1z^{-2},\ n_1z^{-1},\ n_1,\ n_2,\ n_1z,\ n_2z,\ e_1z^2,\ e_2z^2,\ e_3z^2,\
e_1z^3,\ e_2z^3,\ \cdots .
\]
Using this one can calculate in a similar way as in \cite{LM} (using the
boson-fermion correspondence or vertex operator constructions) the wave
function:
\[
\begin{aligned}
\Psi(t,z)=&P(t,z)\exp \left(\sum_{i=1}^n\sum_{j=1}^\infty
t_j^{(i)}z^j E_{ii}\right),\quad\mbox{where }\\
P_{jj}(t,z)=&\frac{\hat\tau(t^{(k)}_\ell-\delta_{kj}(\ell
z^\ell)^{-1})}{\hat\tau(t)},\\
P_{ij}(t,z)=&z^{-1}\frac{\hat\tau_{ij}(t^{(k)}_\ell-\delta_{kj}(\ell
z^\ell)^{-1})}{\hat\tau(t)}
\end{aligned}
\]
and where
\[
\hat\tau(t)=\det
\begin{pmatrix}
n_{11}S_2(t^{(1)})&n_{11}S_1(t^{(1)})&n_{11}&0&n_{12}&0\\
n_{21}S_2(t^{(2)})&n_{21}S_1(t^{(2)})&n_{21}&0&n_{22}&0\\
n_{31}S_2(t^{(3)})&n_{31}S_1(t^{(3)})&n_{31}&0&n_{32}&0\\
n_{11}S_3(t^{(1)})&n_{11}S_2(t^{(1)})&n_{11}S_1(t^{(1)})
&n_{11}&n_{12}S_1(t^{(1)})&n_{12}\\
n_{21}S_3(t^{(2)})&n_{21}S_2(t^{(2)})&n_{21}S_1(t^{(2)})
&n_{21}&n_{22}S_1(t^{(2)})&n_{22}\\
n_{31}S_3(t^{(3)})&n_{31}S_2(t^{(3)})&n_{31}S_1(t^{(3)})
&n_{31}&n_{32}S_1(t^{(3)})&n_{32}\\
\end{pmatrix}.
\]
The functions  $S_i(x)$ are the elementary Schur polynomials:
\[
S_1(x)=x_1,\quad S_2(x)=\frac{x_1^2}{2}+x_2,\quad
S_3(x)=\frac{x_1^3}{6}+x_2x_1+x_3.
\]
The  tau function
$\hat\tau_{ij}(t)$ is up to the sign  sign$(i-j)$ equal to the above
determinant
where we replace the $j$-th row by
\[
\begin{pmatrix}
n_{i1}S_1(t^{(i)})&n_{i1}&0&0&0&0
\end{pmatrix}.
\]
Next we put
 all higher times $t_j^{(i)}$ for
$j>1$ equal to 0 and  write $u_i$ for $t_1^{(i)}$.  Then using the
orthogonality-like  condition
(\ref{N}) of the matrix $N$ to reduce long expressions, the wave function
becomes:
\[
\begin{aligned}
\Psi(u,z)=& \Biggl(I+\frac{1}{\tau(u)}
\sum_{i,j=1}^3
\left[\left(-w^{(3)}_{1}
+w^{(2)}_{1}(u_i+u_j)-w^{(1)}_{1}u_iu_j\right)z^{-1}\right.\\[2mm]
&+\left.\left(w^{(1)}_{1}u_i-w^{(2)}_{1}\right)z^{-2}
\right]n_{i1}n_{j1}E_{ij}
\Biggr)e^{zU},
\end{aligned}
\]
where, for  convenience of
notation, we have  introduced some new "variables"
\[
w^{(k)}_{i}=\frac{1}{k}\sum_{\ell=1}^3 u_\ell^kn_{\ell i}n_{\ell 1},
\]
and where
\[
\tau(u)=w^{(3)}_{1}w^{(1)}_{1}-w^{(2)}_{1}w^{(2)}_{1}.
\]
Note that in this way we also have determined the rotation coefficients
\[
\beta_{ij}=\frac{1}{\tau(u)}\left(-w^{(3)}_{1}
+w^{(2)}_{1}(u_i+u_j)-w^{(1)}_{1}u_iu_j\right)n_{i1}n_{j1},
\]
which is a new solution of order $3$ of the the Darboux--Egoroff equations.

Recall that $\eta=\sum_{i=1}^3 E_{i,4-i}$.
It is now straightforward but tedious to determine the flat coordinates
$x^\alpha$ and the $\phi_\alpha^{(j)}$ for $j>1$. One finds that
for $\ell>0$ and $p=1,2,3$:
\begin{equation}\label{32}
\phi^{(\ell-\mu_p)}_p=\frac{\tau w^{(\ell+2)}_{p}+\tau_1w^{(\ell+1)}_{p}
+\tau_2
w^{(\ell)}_{p}}{2(\ell-1)!\tau}
\end{equation}
and
\begin{equation}
\phi^{(-2-\mu_p)}_p=\delta_{p3},\quad
\phi^{(-1-\mu_p)}_p=-\delta_{p3}\frac{\tau_1}{2\tau},\quad
\phi^{(-\mu_p)}_p=\delta_{p3}\frac{\tau_2}{2\tau},
\end{equation}
where
\[
\begin{aligned}
\tau_1=&w^{(2)}_{1}w^{(3)}_{1}-w^{(1)}_{1}w^{(4)}_{1}\\[2mm]
\tau_2=&w^{(2)}_{1}w^{(4)}_{1}-(w^{(3)}_{1})^2.
\end{aligned}
\]
Note that (\ref{32}) also holds for $p=1$ and $\ell=1,2$, it is easy to verify
that $\phi^{(-1)}_1=\phi^{(0)}_1=0$.
Using (\ref{xione}), one has the following flat coordinates:
\begin{equation}
\begin{aligned}
x^1=&-\frac{\tau_1}{2\tau},\\[2mm]
x^2=&\frac{1}{2\tau}\left(
\tau
w^{(3)}_{2}+\tau_1w^{(2)}_{2}+\tau_2w^{(1)}_{2}\right),\\[2mm]
x^3=&\frac{1}{4\tau}\left(
\tau
w^{(5)}_{1}+\tau_1w^{(4)}_{1}+\tau_2w^{(3)}_{1}\right),
\end{aligned}
\end{equation}
{}From all this it is straightforward to determine $F(u)$, given by
(\ref{fep}):
\[
\begin{aligned}
F&= \frac{\tau_2}{16\tau^2}
\left(\tau w_{1}^{(5)} +\tau_1 w_{1}^{(4)}
+\tau_2 w_{1}^{(3)}\right)\\[2mm]
\ &-\frac{\tau_1}{48\tau^2}
\left(\tau w_{1}^{(6)} +\tau_1 w_{1}^{(5)}
+\tau_2 w_{1}^{(4)}\right)\\[2mm]
\ &-\frac{\tau}{96\tau^2}
\left(\tau w_{1}^{(7)} +\tau_1 w_{1}^{(6)}
+\tau_2 w_{1}^{(5)}\right)\\[2mm]
\ &+\frac{1}{8\tau^2}
\left(\tau w_{2}^{(3)} +\tau_1 w_{2}^{(2)}
+\tau_2 w_{2}^{(1)}\right)
\left(\tau w_{2}^{(4)} +\tau_1 w_{2}^{(3)}
+\tau_2 w_{2}^{(2)}\right).
\end{aligned}
\]
We shall not determine the explicit form of this prepotential in terms of the
canonical coordinates
here, because it is quite long. However, there is a problem even in this
"simple" example. We do not know how to express the cannonical coordinates
$u_i$
in terms of the flat ones, the $x^\alpha$'s and thus cannot express $F$ in
terms
of the flat coordinates. Hence we cannot determine the desired form of $F$.
This can be solved in the simplest example, see \cite{LM}, viz.
the case that $\mu_1=-\mu_n=1$ and all other $\mu_i=0$. This gives a rational
prepotential $F$ (in terms of the flat coordinates).
\vskip 10pt \noindent
{\bf Acknowledgements} \\
H.A. was partially supported by NSF (PHY-9820663).


\end{document}